\documentclass[twocolumn]{aastex631}

\usepackage[colorinlistoftodos]{todonotes}
\usepackage{amssymb}
\usepackage{comment}
\usepackage{multirow,acronym}
\usepackage{natbib}
\usepackage{listings} 
\usepackage{makecell}
\usepackage{booktabs}
\usepackage{amsmath}
\usepackage{subfigure}
\usepackage{color}
\usepackage{xcolor}
\usepackage{hyperref}
\usepackage[T1]{fontenc}
\usepackage{graphicx}
\usepackage{bm}
\usepackage{CJK}
\hypersetup{colorlinks=true, citecolor=blue, 
  linkcolor=cyan, urlcolor=magenta}
\usepackage[linesnumbered,ruled]{algorithm2e}
\newcommand{\code}[1]{\lstinline{#1}}
\newcommand{\kratos}{\code{Kratos}}
\newcommand{\cantera}{\code{Cantera}}
\newcommand{\inter}{\mathrm{int}}
\usepackage{aas_macros}
\usepackage{amsfonts}
\usepackage{float}
\usepackage{amsmath,amssymb}
\usepackage{natbib}    
\usepackage{hyperref} 
\usepackage{graphicx} 
\usepackage{color}
\usepackage{verbatim}
\usepackage{enumitem}
\usepackage{natbib}

\usepackage[linesnumbered,ruled]{algorithm2e}

\newcommand{\proptosim}{\mathrel{\vcenter{
 \offinterlineskip\halign{\hfil$##$\cr
 \propto\cr\noalign{\kern2pt}\sim\cr\noalign{\kern-2pt}}}}}
\renewcommand{\b}[1]{\boldsymbol{#1}}
\newcommand{\unit}[1]{{\rm #1}}

\newcommand{\response}[1]{{\color{blue} #1}}
\newcommand{\mean}[1]{\langle #1\rangle}

\renewcommand{\min}{\mathrm{min}}
\renewcommand{\max}{\mathrm{max}}

\newcommand{\cm}{\unit{cm}}
\newcommand{\m}{\unit{m}}
\newcommand{\g}{\unit{g}}
\newcommand{\G}{\unit{G}}
\newcommand{\K}{\unit{K}}

\newcommand{\pc}{\unit{pc}}

\newcommand{\eV}{\unit{eV}}

\newcommand{\s}{\mathrm{s}}

\newcommand{\B}{\mathbf{B}}     
\newcommand{\E}{\mathbf{E}}     
\newcommand{\J}{\mathbf{J}}     
\renewcommand{\v}{\mathbf{v}}   
\newcommand{\kb}{k_\mathrm{B}}

\renewcommand{\d}{\mathrm{d}}

\newcommand{\e}{\mathrm{e}}

\newcommand{\p}{\mathrm{p}}     

\newcommand{\eff}{\mathrm{eff}}

\newcommand{\ph}{\mathrm{ph}}

\newcommand{\abs} {\mathrm{abs}} 

\renewcommand{\O}{\mathrm{O}}     
\newcommand{\A}{\mathrm{A}}     
\renewcommand{\H}{\mathrm{H}}     
\renewcommand{\P}{\mathrm{P}}     
\newcommand*\chem[1]{\ensuremath{\mathrm{#1}}}

\newcommand{\univec}[1]{\hat{\mathbf{#1}}}
\newcommand{\figdir}{.}
\renewcommand{\response}[1]{#1}

\begin{document}

\title{The \kratos{} Framework for Heterogeneous Astrophysical
  Simulations:\\ Ray Tracing, Reacting Flow and Thermochemistry}

\author[0000-0002-6540-7042]{Lile Wang}
\affil{The Kavli Institute for Astronomy and Astrophysics,
  Peking University, Beijing 100871, China}
\affil{Department of Astronomy, School of Physics, Peking
  University, Beijing 100871, China}

\correspondingauthor{Lile Wang}
\email{lilew@pku.edu.cn}

\begin{abstract}
  Thermochemistry, ray-tracing, and radiation-matter
  interactions are important processes which are
  computationally difficult to model in astrophysical
  simulations, addressed by introducing novel algorithms
  optimized for heterogeneous architectures in the \kratos{}
  framework. Key innovations include a
  stoichiometry-compatible reconstruction scheme for
  consistent chemical species advection, which ensures
  element conservation while avoiding matrix inversions, and
  a LU decomposition method specifically designed for
  multi-thread parallelization in order to solve stiff
  thermochemical ordinary differential equations with high
  efficiency. The framework also implements efficient
  ray-tracing techniques for radiation-matter
  interactions. Various verification tests, spanning from
  chemical advection, combustion, Str\"omgren spheres, and
  detonation dynamics, are conducted to demonstrate the
  accuracy and robustness of \kratos{}, with results closely
  matching semi-analytic solutions and benchmarks such as
  \cantera{} and the Shock and Detonation Toolbox.  The
  modular design and performance optimizations position it
  as a versatile tool for studying coupled microphysical
  processes in the diverse environments of contemporary
  astrophysical studies.
\end{abstract}


\section{Introduction}
\label{sec:intro}

Astrophysical reacting flows are among the most complex and
fascinating phenomena in the universe, involving a wide
range of physical processes such as hydrodynamics,
magnetohydrodynamics (MHD), thermochemistry, and
radiation. These processes are intricately coupled and play
crucial roles in various astrophysical phenomena, including
planet formation, star formation, stellar evolution,
supernova explosions, and the formation of galaxies. To gain
a deeper understanding of these phenomena, numerical
simulations have become an essential tool, allowing
researchers to model and analyze the underlying mechanisms
in detail.

Consistent microphysics calculations, including
thermochemistry and radiation, are emerging as necessities
for numerical simulations of astrophysical processes
\citep[e.g.][] {2024arXiv241112491Z}. Thermochemistry, in
particular, is a critical component that governs the
chemical reactions and energy release in astrophysical
environments. However, thermochemical calculations are often
computationally expensive due to the stiffness of the
thermochemical systems, which makes them challenging to
integrate into dynamic simulations. As a result, most
previous works have either focused on static grids or
neglected the consistency between fluid mechanics and
thermochemistry . This limitation has hindered the
development of comprehensive models that can accurately
capture the full complexity of astrophysical reacting flows.

Existing computational codes that claim to be at the
forefront of research often face several challenges. First,
many codes have restricted availability or are subject to
copyright limitations, which restricts their widespread use
and collaboration opportunities. Second, some codes lack the
capability to efficiently integrate with hydrodynamic
calculations, often due to the computational inefficiencies
associated with thermochemical calculations. Finally, the
reliability and accuracy of these codes are not always
well-controlled, which can lead to uncertainties in the
simulation results .

To address these challenges, the \kratos{} code based on
heterogeneous devices, especially GPUs (graphical processing
units), aims to provide a comprehensive and efficient
solution for simulating astrophysical reacting flows. By
integrating consistent microphysics calculations with
hydrodynamics, the system seeks to overcome the limitations
of previous approaches and enable more accurate and reliable
simulations of astrophysical phenomena . This work is
expected to contribute significantly to the field by
providing a powerful tool for researchers to explore the
complex interactions between different physical processes in
astrophysical environments.  The method described in this
paper has actually been adopted in several astrophysical
works already. An earlier version of the GPU-based
thermochemistry module developed by the author, although
based on the hydrodynamics and MHD of \code{Athena++}
\citep{stone_athena_2020}, have been applied in various
astrophysical scenarios, including protoplanetary disk
photoevaporation \citep{2017ApJ...847...11W} and magnetized
disk winds \citep{2019ApJ...874...90W, 2019ApJ...885...36H,
  2020ApJ...904L..27N, 2023NatAs...7..905F}, exoplanetary
atmospheres \citep{2018ApJ...860..175W, 2019ApJ...873L...1W,
  2021ApJ...914...98W, 2021ApJ...914...99W}, and
interstellar media \citep{2024ApJ...973...37Y}. Implementing
the microphysics modules on \kratos{}, a platform with
optimized algorithms and procedures intrinsically designed
for GPUs, offers a high-performance platform including
co-evolved with real-time microphysics, which are crucial to
various incoming comprehensive simulations that aim at the
consistency of astrophysical processes studies.

This paper is structured as follows. \S\ref{sec:method-chem}
elaborates the methods for thermochemistry specifically
adapted for GPUs, focusing on the conservation laws and the
parallelization of algorithms. \S\ref{sec:method-rt}
describes the GPU-optimized ray tracing method, including
the geometric calculations on the structured mesh of
\kratos{} and the details of radiation-materials
interactions based on the structured
mesh. \S\ref{sec:verify} exhibits verifications and tests
regarding the microphysics modules, including
thermochemistry alone, radiation alone, and the tests that
comprehensively involve the interactions of fluids,
chemicals, and radiation. The implementation and results of
these methods are summarized in \S\ref{sec:summary}, along
with discussion about prospective future improvements.

\section{Thermochemistry and Reacting Flows}
\label{sec:method-chem}

The reacting flow module elaborated in this paper is part of
\kratos{}, a grid-based heterogeneous system for
astrophysical simulations \citep{2025arXiv250102317W}. The
flexiblity of the \kratos{} framework and its hydrodynamic
module allows straightforward implementations of consistent
advection schemes for chemical species
(\S\ref{sec:method-cons-advec}), which is the foundation of
all procedures for reacting flow computations. Similar to
the hydrodynamics module described in
\citep{2025arXiv250102317W}, the thermochemistry module also
has three modes in terms of floating point precision,
including a full single-precision mode, a full
double-precision mode, and a mixed-precision mode that use
double-precision operations only when updating the conserved
quantities (the final abundance of chemical species at the
end of each thermochemistry cycle).  It is noted that
\kratos{} allows the user to select the subset of chemical
species and reactions involved, simply by assigning the list
of the chemical species and the file of reaction database in
the input file.

\subsection{Conservation of Chemicals in Fluid Advection}
\label{sec:method-cons-advec}

A wide range of applications within the \kratos{} framework,
including all computations discussed in this work, rely on
the higher order schemes for reconstruction (e.g., piecewise
linear method; PLM) and a Riemann solver (e.g., HLLC) for
hydrodynamic fluxes. This combination achieves a balance
between computational efficiency and numerical accuracy,
making it suitable for simulating complex astrophysical
reacting flows. In this section, we focus on the advection
of chemical species, which is a critical component of the
framework, as it ensures the conservation of chemical
elements while maintaining numerical stability.

The advection scheme in \kratos{} is implemented as a
Godunov solver, which consists of three primary steps: (1)
reconstruction of variable values at cell interfaces, (2)
calculation of fluxes using a Riemann solver, and (3)
updating variables by adding the divergence of
fluxes. Conservation of chemical elements is inherently
guaranteed in the third step, provided that the first two
steps are implemented correctly. This is particularly
important in astrophysical simulations, where strict
conservation of chemical abundances is required unless
nuclear reactions are explicitly involved.  To illustrate
the methodology, a one-dimensional problems is considered,
though the schemes can be straightforwardly extended to
three dimensions. The flux of the sth chemical species at
the $i-1/2$ surface (left surface of the $i$th cell) is
given by:
\begin{equation}
  \label{eq:method-flux-sp}
  \mathcal{F}^s_{i-1/2} = x^s_{i-1/2}
  \mathcal{F}^\rho_{i-1/2}\ ; \quad
  x^s_{i-1/2} \equiv \left[\dfrac{n^s}{\rho} \right]_{i-1/2}\ ,
\end{equation}
where $[n^s/\rho]_{i-1/2}$ represents the ratio of the
number density of the sth species to the mass density at the
$i-1/2$ surface. Here, $F^s_{i- 1/2}$ and
$\mathcal{F}^\rho_{i-1/2}$ denote the fluxes of the $s$th
species in terms of molecule numbers and mass density,
respectively. This formulation ensures that the advection of
chemical species is consistent with the underlying
hydrodynamics, thereby preserving the integrity of chemical
element abundances throughout the simulation domain.

It is straightforward to formulate the element conservation
condition for chemical species fluxes at cell
interfaces. For each element indexed by $\nu$, the
conservation relationship can be expressed as:
\begin{equation}
  \label{eq:method-flux-elem-cond}
  \sum_{s} \mathcal{N}_{\nu, s}  \mathcal{F}^s_{i- 1/2}
  = \dfrac{1}{\alpha} X_\nu \mathcal{F}^\rho_{i- 1/2},
\end{equation}
where $s$ iterates over all chemical species,
$\mathcal{N}_{\nu, s}$ denotes the number of atoms of the
$\nu^{\text{th}}$ element in the $s^{\text{th}}$ species,
$X_\nu$ represents the relative nuclear abundance of the
$\nu^{\text{th}}$ element, and
$\alpha \equiv \sum_{\nu} X_\nu m_\nu$ defines the mean
\textit{atomic} mass. Eq.~\eqref{eq:method-flux-elem-cond}
constitutes an underdetermined system when the number of
species exceeds the number of elements. This conservation
condition remains mathematically well-posed only under the
assumption of constant elemental abundances, formally
expressed as,
\begin{equation}
  \label{eq:method-const-elem}
  \sum_{s} \mathcal{N}_{\nu,s}n^s_i = \dfrac{1}{\alpha}X_\nu
  \rho_i,\quad \forall\ i. 
\end{equation}

A critical examination reveals that directly employing
absolute or relative abundances in numerical schemes while
allowing independent reconstruction and Riemann solver for
individual species proves fundamentally flawed. This
approach fails because higher-order reconstruction methods
(e.g., piecewise linear methods) employ nonlinear
transformations during interface value
reconstruction. Consequently, the element conservation
requirement stipulated by
eq.~\eqref{eq:method-flux-elem-cond} becomes
systematically violated in such implementations. The
nonlinear nature of these reconstruction operators disrupts
the linear proportionality between species fluxes and
density fluxes that element conservation demands.

\subsubsection{Consistent Multifluid Advection (CMA)}
\label{sec:method-cma}

To maintain accurate elemental abundances in multispecies
fluid simulations, the concept of Consistent Multifluid
Advection (CMA) was pioneered by \citet{1999A&A...342..179P}
and significantly advanced by
\citet{2010MNRAS.404....2G}. Although possible if users
desire, it is emphasized that CMA is \textit{not} included
in \kratos{} by default, and the following descriptions are
presented only for logical completeness.

The CMA approach modifies the reconstructed relative species
abundances $\{\tilde{x}^s_{i-1/2}\}$ through element
conservation constraints. The corrected interface abundances
$\{x^s_{i-1/2}\}$ are computed via:
\begin{equation}
  \label{eq:method-cma-def-eta}
  x^s_{i-1/2} = \sum_{\nu} 
  \dfrac{\mathcal{N}_{\nu,s}}{\mathcal{N}_{\mathrm{tot},s}} \eta_\nu 
  \tilde{x}^s_{i-1/2},
\end{equation}
where
$\mathcal{N}_{\mathrm{tot},s} \equiv \sum_{\nu}
\mathcal{N}_{\nu,s}$ denotes the total nuclei count per
molecule of species $s$. The correction coefficients
$\{\eta_\nu\}$ are determined by solving the linear system:
\begin{equation}
  \label{eq:method-cma-eta-sys}
  \begin{split}
    \sum_{s} \mathcal{N}_{\mu,s} x^s_{i-1/2}
    & \equiv \sum_{\nu} \mathcal{M}_{\mu,\nu} \eta_\nu =
      \dfrac{X_\mu}{\alpha}, \\ 
    \mathcal{M}_{\mu,\nu}
    & \equiv \sum_{s} \mathcal{N}_{\mu,s}
      \dfrac{\mathcal{N}_{\nu,s}}{\mathcal{N}_{\mathrm{tot},s}}
      \tilde{x}^s_{i-1/2}\ ,
  \end{split}
\end{equation}
where $\mathcal{M}_{\mu,\nu}$ constitutes an element
coupling matrix. This $O(N_{\rm elem})$ linear system
requires independent solution at each cell interface
$(i-1/2)$, whose computational costs are expensive at the
order of $O(N_{\rm elem}^3)$. More crucially, the modified
reconstruction $\{x^s_{i-1/2}\}$ preserves the spatial
accuracy order of the original scheme applied to
$\{\tilde{x}^s_{i-1/2}\}$.

\subsubsection{Donor-cell Reconstruction:\\
  Conserved, but Lowest Order}
\label{sec:method-donor-cell}

A physically intuitive approach to determining
$\{x^s_{i-1/2}\}$ emerges from Riemann solver behavior. The
Riemann solvers naturally capture contact discontinuities
where no species mixing occurs across wave surfaces--a
property enforced by the Courant-Friedrichs-Lewy (CFL)
condition. This implies that the relative species abundance
at cell interfaces should inherit values from the upwind
cell, determined by the contact surface's motion
direction. This motivates a donor-cell-type scheme
formulated as:
\begin{equation}
  \label{eq:method-donor-cell-recons}
  x^s_{i-1/2} 
  = \left(\dfrac{n^s_{i-1}}{\rho_{i-1}}\right)
  \Theta(\mathcal{F}^\rho_{i-1/2})
  + \left(\dfrac{n^s_i}{\rho_i}\right)
  \Theta(-\mathcal{F}^\rho_{i-1/2}),
\end{equation}
where $\Theta(x)$ represents the Heaviside step function
($0$ if $x\leq 0$ and $1$ otherwise).

Under the global element abundance constraint
(eq.~\ref{eq:method-const-elem}), this formulation
inherently satisfies the flux conservation condition in
eq.~\eqref{eq:method-flux-elem-cond}. Key advantages of this
scheme include the minimal computational overhead through
direct upwind selection, and the compatibility with
spatially varying element abundances (unlike the CMA
method). Nonetheless, significant diffusivity stems from
strict proportionality between species fluxes and the upwind
values (eq.~\ref{eq:method-flux-sp}), which remains as an
apparent issue to be resolved.

\subsubsection{Stoichiometry-compatible Reconstruction:\\
  Matrix Inversion Avoided for Higher Order}
\label{sec:method-storec}

\response{ To match the contemporary reconstruction methods
  with higher spatial order of accuracy, an alternative
  higher-order scheme is implemented in \kratos{}, which
  avoids computational bottlenecks associated with matrix
  inversion in CMA while surpassing the spatial accuracy of
  donor-cell methods}. Central to this approach is the
stoichiometric space $\mathcal{S}$, defined as the null
space of the element-species matrix
$\mathcal{N}_{\nu,s}$. By construction, any variation
$\Delta n^s \in \mathcal{S}$ preserves elemental abundances,
\begin{equation}
    n^s \rightarrow (n^s + \Delta n^s)\ ,\ 
    \text{with} \  \sum_s \mathcal{N}_{\nu,s}\Delta
    n^s = 0\ ,\ \forall \nu\ .
\end{equation}
Let $\mathcal{P}$ denote the projection operator associated
with $\mathcal{N}_{\nu,s}$, which can be constructed during
initialization through singular value decomposition (SVD) of
the stoichiometric matrix,
\begin{equation}
    \mathcal{N}_{\nu,s} = \mathbf{U}\Sigma\mathbf{V}^\top
    \Rightarrow \mathcal{P} =
    \mathbf{V}\mathbf{V}^\top\ , 
\end{equation}
where $\mathbf{V}$ and $\mathbf{U}$ contains the right and
left singular vectors spanning the right and left spaces of
$\mathcal{S}$, and $\Sigma$ is the diagonal matrix for
singular values. The SVD procedure are included in
\kratos{}, which typically accounts for a negligible
fraction of total runtime in practical simulations, as it
only needs to be conducted once at the beginning of each
run.

The interface states can be expressed through generalized
reconstruction:
\begin{equation}
  \label{eq:method-recon-general}
  \begin{split}
    x^s_{i-1/2,l}  & = \mean{x}^s_{i-1} + \chi^s_{i-1/2,l} \\
    x^s_{i-1/2,r} & = \mean{x}^s_i + \chi^s_{i-1/2,r}
  \end{split}
\end{equation}
where $\chi^s_{i-1/2,l/r}$ contains higher-order correction
terms. For instance, in the piecewise linear method (PLM)
reconstruction scheme adopted in \kratos{} by default, these
terms take the form,
\begin{equation}
  \label{eq:method-plm-general}
  \begin{split}
    \chi^s_{i-1/2,l}
    & = \phi\left(\partial x^s_{i-1,f},\partial
      x^s_{i-1,b},c_{i-1,f},c_{i-1,b}\right) \\ 
    & \quad \times (\xi_{i-1/2} - \mean{\xi}_{i-1}) \\
    \chi^s_{i-1/2,r}
    & = \phi\left(\partial x^s_{i,f},\partial
      x^s_{i,b},c_{i,f},c_{i,b}\right) \\ 
    & \quad \times (\xi_{i-1/2} - \mean{\xi}_i)\ ,
  \end{split}
\end{equation}
where $\xi$ denotes spatial coordinate, $\xi_{i-1/2}$ the
interface position, and $\mean{\xi}_i$ the cell-centered
coordinate. The gradient terms and geometric coefficients
are defined as,
\begin{align}
  \label{eq:method-sto-plm-def}
    \partial x^s_{i-1,f}
    & \equiv \dfrac{\mean{x}^s_i -
      \mean{x}^s_{i-1}}{\mean{\xi}_i - \mean{\xi}_{i-1}},
      \   
    \partial x^s_{i-1,b} \equiv \dfrac{\mean{x}^s_{i-1} -
      \mean{x}^s_{i-2}}{\mean{\xi}_{i-1} -
      \mean{\xi}_{i-2}}\ ;
  \nonumber  \\ 
    c_{i-1,f}
    & \equiv \dfrac{\mean{\xi}_i -
      \mean{\xi}_{i-1}}{\xi_{i-1/2} - \mean{\xi}_{i-2}},
      \   
    c_{i-1,b} \equiv \dfrac{\mean{\xi}_{i-1} -
      \mean{\xi}_{i-2}}{\mean{\xi}_{i-1} - \xi_{i-3/2}}\ .
\end{align}
Mirror expressions hold for right-side coefficients
$c_{i,f}$, $c_{i,b}$. The slope limiter $\phi$ implements
the modified van Leer formulation from
\citet{2014JCoPh.270..784M},
ensuring monotonicity preservation through:
\begin{equation}
    \phi(a,b,c,d) = \begin{cases} 
        0 &  ab \leq 0\ ; \\
        \min\left[\dfrac{2|a|}{|a|+|b|}, 
      \dfrac{2c}{c+d}\right]\text{sgn}(a), & ab > 0\ .
    \end{cases}
\end{equation}
The nonlinear nature of slope limiters introduces
fundamental challenges for chemical conservation: nonlinear
reconstruction operators generally violate the
proportionality required by
eq.~\eqref{eq:method-flux-elem-cond}. This issue is
resolved through stoichiometric space projection,
reformulating the interface states as:
\begin{equation}
  \label{eq:method-sto-plm-proj}
  x^s_{i-1/2,l} = \mean{x}^s_{i-1} + \sum_r \mathcal{P}^s_r
  \chi^r_{i-1/2,l}, 
\end{equation}
with an analogous expression for $x^s_{i-1/2,r}$. The full
interface reconstruction then follows:
\begin{equation}
  \label{eq:method-sto-plm-recon}
  x^s_{i-1/2} = x^s_{i-1/2,l}
  \Theta(\mathcal{F}^\rho_{i-1/2}) + x^s_{i-1/2,r}
  \Theta(-\mathcal{F}^\rho_{i-1/2}). 
\end{equation}
This formulation exhibits features accuracy to arbitrary
spatial order, as the correction terms $\chi$ are not
limited to first order in
eq.~\eqref{eq:method-recon-general}. Non-zero projections
$\mathcal{P}\chi$ introduce spatially varying corrections
while preserving stoichiometric constraints, achieving
formal second-order accuracy through the slope-limited
derivatives. Note also that the cost of calculating
$\mathcal{P}\chi$ is at the order of $O(N^2)$, which is
generally much quicker and free from numercial instabilities
that are involved in the matrix inversion procedures.

\subsection{Thermochemical Evolution}
\label{sec:method-algo-cpu}

The thermochemical evolution in each computational cell is
governed by a coupled system of stiff ordinary differential
equations (ODEs) for chemical species densities $\{n^i\}$
and internal energy density $\epsilon$:
\begin{equation}
  \label{eq:method-ode-thermochem}
  \begin{split}
    \dfrac{\d n^i}{\d t}
    & = \sum_{j,k} \mathcal{A}^i_{\;jk} n^j n^k + \sum_j
      \mathcal{B}^i_{\;j} n^j, \\ 
    \dfrac{\d \epsilon}{\d t}
    & = \Gamma(\{n^i\}, T) - \Lambda(\{n^i\}, T)\ ,
  \end{split}
\end{equation}
where $\mathcal{A}^i_{\;jk}$ represents rate coefficients
for binary reactions (formation or destruction of species
$i$ via interactions between $j$ \& $k$),
$\mathcal{B}^i_{\;j}$ captures unary processes affecting
species $i$ (especially photoionization, radioactive decay,
etc.), and $\Gamma$ and $\Lambda$ denote volumetric heating
and cooling rates, respectively. All reaction coefficients
could be functions of the temperature $T$, which is
thermodynamically coupled through
$T = {\epsilon}/c_V(T, \{n^i\})$; $c_V$ is the
constant-volume heat capacity derived from the equation of
state, as a function of the temperature itself and the
abundances of all species $\{n^i\}$. The use of $c_V$
(rather than $c_P$) reflects the operator-split approach
where thermochemistry evolves independently from
hydrodynamic work terms, as the hydrodynamic steps already
resolve the contribution of volumetric work on its own.

Key numerical challenges typically arise from the stiffness,
reflecting widely separated timescales between chemical
processes (e.g., fast ionizations versus slow molecular
reactions).  Non-linear dependence could also add to the
difficulties, including the temperature dependence of rate
coefficients $\mathcal{A}^i_{\;jk}(T)$,
$\mathcal{B}^i_{\;j}(T)$ combined with thermal feedback
through $\Gamma(T)$ and $\Lambda(T)$. In what follows, the
numerical methodology is detailed for robust integration of
this ODE system, emphasizing stability preservation and
computational efficiency.

\subsubsection{Reaction Rates and Jacobian Matrices}
\label{sec:method-rate-jac}

The integration of thermochemical ODEs necessitates careful
evaluation of reaction rate coefficients, which are
categorized roughtly into three distinct types:
\begin{enumerate}
\item Two-body reactions with modified Arrhenius
  rate coefficients,
    \begin{equation}
        k(T) = k_0 \left(\dfrac{T}{T_0}\right)^\beta
        \exp\left(-\dfrac{T_0}{T}\right) ,
    \end{equation}
    where $(k_0, T_0, \beta)$ are parameters from
    astrochemical databases like
    \citet{UMIST2013}. Associated heating/cooling rates
    follow similar temperature dependence.
  \item Photoreactions governed by effective radiation flux
    $F_{\mathrm{ph,eff}}$, 
    \begin{equation}
        \zeta = \sum_i n^i \sigma^i_{\mathrm{abs}}
        F_{\mathrm{ph,eff}}(\lambda^{-1}_{\mathrm{abs}})\ ,, 
    \end{equation}
    where $\lambda^{-1}_{\mathrm{abs}} = \sum_i n^i
    \sigma^i_{\mathrm{abs}}$ is the inverse absorption
    length. The flux $F_{\mathrm{ph,eff}}$ self-consistently
    adapts to evolving chemical abundances (see also
    \S\ref{sec:method-rad-reac}). 
  \item Tabulated reactions with numerically interpolated
    rates, where precomputed tables are read in to account
    for complex temperature dependence and local conditions
    (e.g., photon escape probabilities). At the beginning of
    the thermochemical computation, multi-dimensional tables
    are reduced to one-dimensional in $T$ per cell, assuming
    fixed total mass density and elemental abundances during
    thermochemical steps.
\end{enumerate}

For efficient solution of stiff ODEs using semi-implicit
methods, the elements of Jacobian matrices,
\begin{equation}
  \label{eq:method-jacobian}
  \mathbf{J} \equiv \left[
    \begin{matrix}
      \partial \dot{n}^i/\partial n^j,
      & \partial \dot{n}^i/\partial \epsilon \\
      \partial \dot{\epsilon}/\partial n^j,
      & \partial \dot{\epsilon}/\partial \epsilon \\
    \end{matrix}
  \right]\ ,
\end{equation}
are estimated analytically through (1) mass action derivatives:
\begin{equation}
  \dfrac{\partial (\mathcal{A}^i_{jk}n^jn^k)}{\partial n^m} = 
  \begin{cases}
    2\mathcal{A}^i_{mm}n^m & (m = j = k) \\
    \mathcal{A}^i_{jk}n^k & (m = j \neq k) \\
    \mathcal{A}^i_{jk}n^j & (m = k \neq j)
  \end{cases}, 
\end{equation}
(2) thermal coupling terms ( with analytical
$\partial k/\partial T$ for Arrhenius and tabulated rates;
the heat capacity derivative $\partial c_V/\partial n^i$ is
typically neglected in the approximations for $\mathbf{J}$),
\begin{equation}
  \dfrac{\partial k}{\partial \epsilon} = \dfrac{\partial
    k}{\partial T}\dfrac{\partial T}{\partial \epsilon} =
  \dfrac{1}{c_V}\dfrac{\partial k}{\partial T}\ , 
\end{equation}
and (3) radiation flux derivatives:
\begin{equation}
  \dfrac{\partial F_{\mathrm{ph,eff}}}{\partial n^i}
  \approx \sigma^i_{\mathrm{abs}}
  \left(\dfrac{\partial F_{\mathrm{ph,eff}}}{\partial
    \lambda^{-1}_{\mathrm{abs}}}\right)
  \mathrm{min}\left\{1,
    \dfrac{n^i\sigma^i_{\mathrm{abs}}}{\lambda^{-1}_{\mathrm{abs}}}
  \right\}\ , 
\end{equation}
where the cutoff factor ensures dominant absorbers drive the
derivative calculation. This hybrid analytical-numerical
approach maintains $\mathcal{O}(N_{\rm reac})$ complexity
with respect to reaction number $N_{\rm reac}$, while
capturing essential Jacobian features for stable
integration.

\subsubsection{Pre-processing for Thermochemical
  Computations}
\label{sec:method-pre-proc}

The ODEs are solved as initial value problem, with the
initial values taken from the step of fluid mechanical
evolution that has just been finished. Note that \kratos{}
uses total energy density for its Godunov solver; the
thermal energy density $\epsilon$ is obtained by subtracting
the kinetic energy density (plus magnetic energy density in
magnetohydrodynamic simulations) from the total.

Post-advection chemical abundances $\{n^s\}$ require
validation against numerical floors. If density floors
apply, the abundances of species have to be rescaled.  For
temperature $T = \epsilon/c_V$, the check against effective
tempreature limits also applies to get rid of the
potentially unphysical temperatures introduced by pressure
floors in the hydrodynamic procedures. Prior to ODE
integration, each cell undergoes:
\begin{itemize}
\item Rate table generation: Construct temperature-dependent
  rate interpolation tables, whose selection of temperature
  points should match the reaction scenarios.
\item Dominant absorber identification: Determine species
  $s^*$ maximizing photoabsorption contribution, which
  governs radiation flux derivative approximations during
  Jacobian construction
\end{itemize}
These preprocessing steps ensure thermodynamic consistency
between advected and reacted states, and enhances numerical
stability and computational efficiency through localized
rate approximations.

\subsubsection{Post-processing}
\label{sec:method-post-proc}

Prior to updating the thermochemically evolved abundances
and thermal energy density for subsequent fluid mechanical
steps, several post-processing procedures must be applied to
each computational cell. These safeguards prevent unphysical
outcomes from contaminating the entire simulation.

Frequent minor numerical artifacts require systematic
correction, which mainly involved floating-point precision
artifacts (particularly pronounced in GPU-accelerated
computations using single-precision arithmetic), and
occasional negative abundances due to numerical dispersions.
The regularization pipeline implements three-stage
correction:
\begin{enumerate}
\item If some species have negative abundances, their
  abundances are set to zero;
\item Abundances of elements are calculated for the initial
  and final states of thermochemical evolution. In the final
  state, the abundances in the single-element species are
  adjusted to compensate the differences. This step corrects
  the violation in element abundances brought by step 1;
\item If step 2 causes negative abundances in some
  single-element species, the most abundant compound species
  is disintegrated into elements just sufficiently to
  correct this. As the thermochemical reaction rates
  generally follow the mass-action law, this will cause
  minimal impacts on the behaviors of subsequent processes. 
\end{enumerate}
For cells where the ODE solver reports significant
convergence failures in thermochemical calculations, or
where results exhibit non-physical characteristics (e.g.,
\code{NaN} values, significant negative abundances, or
substantial elemental conservation violations), a retry
procedure is initiated using identical initial conditions
but with reduced initial time step size and enhanced step
size control parameters. Persistent failures trigger either
reversion to pre-evolution values (as specified in
initialization parameters), or a full simulation
termination.

    
    

\subsection{GPU-Optimized Algorithm}
\label{sec:method-ode-gpu}

The computationally intensive nature of thermochemical ODE
systems motivates implementation optimizations for modern
heterogeneous architectures. While these ODE solutions
remain expensive per-cell, their inherent embarrassingly
parallel structure enables efficient acceleration.  Each
computational cell's thermochemical evolution is assigned to
a dedicated thread block, leveraging the massive parallelism
ability of heterogeneous devices.  Because of the special
heterogeneous architectures, the algorithms should be
adjusted to optimize their performances, especially by
utilizing the single-instruction multiple-threads (SIMT)
schemes.

\subsubsection{Reaction Rates and Jacobian Matrices}
\label{sec:method-ode-gpu-rate-jac}

The evaluation of reaction rates and Jacobian matrices
employs SIMT-optimized kernel design. Each thread computes
one reaction's contribution to the overall reaction rate
vector, and thread blocks process reactions in batches
matching warp size (typically 32 or 64 reactions per
batch),
\begin{equation}
    \text{Batch Size} = \min
    \left\{S_{\rm warp}, N_{\rm reac} - k S_{\rm warp} \right\}\ ,
\end{equation}
where $S_{\rm warp}$ is the warp size, and $k$ represents
completed batches.  Excess threads remain inactive during
final partial batches to avoid inconsistent calculations.
Shared memory buffers are applied to eliminate the
high-latency global memory access during rate accumulations.
Results are accumulated by each thread via atomic operations
(using the \code{atomicAdd} interface) to prevent race
conditions.

\subsubsection{Solving Linear Systems: \\
  LU Decomposition with SIMT}
\label{sec:method-ode-gpu-lu}

The computational bottleneck in semi-implicit ODE solvers
emerges from solving numerous small-scale linear systems,
typically involving $10-30$ chemical species per cell,
across massive ensembles of computational
cells. Conventional GPU-optimized linear algebra routines
prove suboptimal for this specific workload pattern, as they
primarily target individual large matrices, dense or sparse,
rather than concurrent batches of small systems. This
architectural mismatch necessitates custom implementation
strategies tailored to the single-instruction
multiple-threads (SIMT) paradigm.

\begin{algorithm}
  \SetKwInOut{Input}{Data}
  \SetKwInOut{Output}{Result}
  \Input{Matrix to be decomposed $a_{ij}$ (size $N\times N$)}
  \Output{LU-decomposed matrix stored in $a_{ij}$;
    Permutation vector $p_i$.} 
  $i\leftarrow \mathrm{thread\ index}\ (0\leq i \leq N -1
  )$\ ;
  
  $v_i \leftarrow$ maximum absolute value in the $i$th row of
  $a_{ij}$
  
  \For{$j \leftarrow 0$ \KwTo $N-1$ } { \tcc{Compute U and
      unrescaled L} \lFor{$k \leftarrow 0$ \KwTo
      $\min\{i-1,j-1\}$ }
    {$a_{ij} \leftarrow a_{ij} - a_{ik}a_{kj}$}
    \tcc{Pivoting: Let diagonals hold greatest possible
      absolute value }
    {synchronize threads;}
    
    \lIf{$i \geq j$}
    {$t_i \leftarrow |a_{ij}| / v_i$}
    {$i_\max\leftarrow$ index for the maximum $t_i$ with
      $i\geq j$;}
    
    {synchronize threads;}
    
    \lIf{$j\neq i_\max$}{swap($a_{i_\max,i}$,$a_{ji}$)}
    \lIf{$i \mathrm{\ is\ } 0$ }{swap($v_{i_\max}$,$v_j$),
      $p_i\leftarrow i_\max$} 
    
    \tcc{Rescale L}

    synchronize threads;
    
    \lIf{$j<i$}{ $a_{ij}\leftarrow a_{ij} / a_{jj}$ }
  }

  \caption{Parallel LU decomposition of matrices on
    multi-thread devices 
    \label{algo:lu-decomp}}
\end{algorithm}

The implementation in \kratos{} employs Crout's algorithm
with row pivoting for LU decomposition
\citep{2002nrca.book.....P}, specifically adapted for GPU
architectures. As detailed in
Algorithm~\ref{algo:lu-decomp}, the computation follows a
thread-parallel approach where all threads execute identical
instructions while maintaining individual thread indices
$i$. Each thread operates on shared matrix $a_{ij}$ and
vector $v_i$ data structures, with explicit synchronization
barriers enforced at critical computation stages (lines 5,
8, and 11) to resolve data dependencies within matrix
groups. Following successful LU decomposition with
permutation tracking, the subsequent backward substitution
phases implement standard parallel reduction patterns across
thread blocks. The SIMT architecture proves particularly
effective for this workload, as the uniform instruction flow
across threads minimizes warp divergence while maximizing
memory coalescing during matrix element access.

\section{Ray Tracing Module}
\label{sec:method-rt}

Radiation fundamentally governs the thermal and chemical
states of astrophysical fluids in many cases, serving as
critical input to thermochemical calculations. The
\response{ray tracing module} in \kratos{} applies direct
ray-tracing calculations for radiation fields in which
scattering is of secondary importance. Although the
extension to a full Monte-Carlo radiation scheme seems
straightforward, \response{the implementation have been
  discussed by \citet{2025arXiv251201283Y} in details.}

\subsection{Ray tracing on Structured Meshes}
\label{sec:method-direct-rt}

\begin{figure}
  \centering
  \includegraphics[width=2.5in, keepaspectratio]
  {\figdir/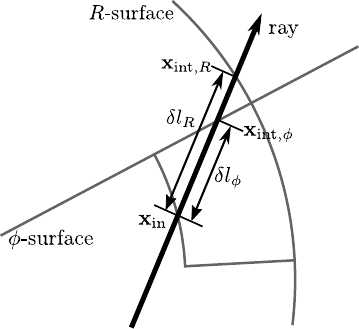} 
  \caption{Example of comparing $\delta l_\phi$ and
    $\delta l_R$ in cylindrical mesh, for the ray passing
    through a cell. This is a top-down view with $z$-related
    quantities omitted for clarity. Grey lines and curves
    indicate the excerpts of coordinate surfaces and
    boundaries of the cell concerned.}
  \label{fig:off_center_cyl}
\end{figure}

\begin{figure}
  \centering
  \includegraphics[width=0.85\linewidth, keepaspectratio]
  {\figdir/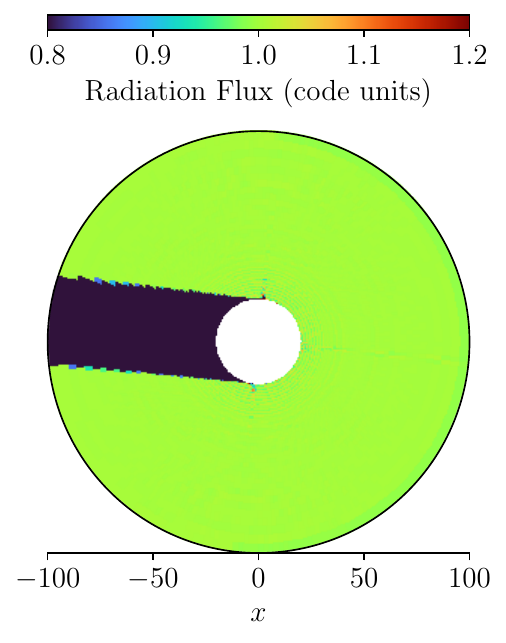} 
  \caption{Test example of ray tracing on a spherical polar
    grid, with the radiation flux (at unitary intensity in
    code units) incidented at a $\theta = \arctan(0.1)$
    angle with the horizontal axis
    (\S\ref{sec:method-direct-rt}). The shadow behind the
    central blank zone can be clearly observed. }
  \label{fig:rad_tst_sph}
\end{figure}

The determination of photon propagation paths is the central
geometric component in \response{ray-tracing
  calculations}. For consistent three-dimensional ray
tracing, precise identification of a ray's entry and exit
points across computational cells is essential. Each cell
interface is mathematically described by six surfaces
$\{S_i(\mathbf{x}) = 0\}$ where $i = 1,\ldots,6$, with
$\mathbf{x}$ denoting spatial coordinates. Given an initial
penetration point $\mathbf{x}_0$ and propagation direction
$\hat{r}$ (where $|\hat{r}| = 1$), the ray trajectory
follows the parametric form
$\mathbf{x}(t) = \mathbf{x}_0 + \hat{r} \delta l $ for
$\delta l \geq 0$.  The set of six equations,
\begin{equation}
  \label{eq:method-rt-geometry}
  S_i(\mathbf{x}_0 + \delta l_i \hat{r} ) = 0\ ;\quad i=1\
  ,\cdots \ ,6\ ,
\end{equation}
are solved for $\{\delta l_i\}$.  In Cartesian coordinates:
\begin{equation}
  \label{eq:cart-dl}
  \begin{split}
    0 = x_{d,i} \delta l_i + ( x_{{\rm in},i} - x_{\inter,i} )
    \quad(i = 1,\ 2,\ 3\ \text{for}\ x,\ y,\ z)\ ,
  \end{split}
\end{equation}
where variables with subscript ``in'' stands for the point
at which the ray gets into the cell, the subscript ``$d$''
indicates the {\it Cartesian} components of the direction
vector $\hat{r}$, and the subscript ``int'' denotes the
coordinate value of the surface to intercept the current
ray. In cylindrical
coordinates:
\begin{equation}
  \label{eq:cyl-dl}
  \begin{split}
    0 & = (x_d^2 + {y}_d^2) \delta l_R^2
    + 2 (x_{\rm in} {x}_d + y_{\rm in} {y}_d) \delta l_R
    + ( R_{\rm in}^2 - R_\inter^2 )\ ;
    \\
    0 & = ({x}_d \tan\phi_\inter - {y}_d)
    \delta l_\phi + (x_{\rm in} \tan\phi_\inter - y_{\rm in})\ ;
    \\
    0 & = {z}_d \delta l_z + ( z_{\rm in} - z_\inter )\ .
  \end{split}
\end{equation}
In spherical polar coordinates:
\begin{equation}
  \label{eq:sph-dl}
  \begin{split}
    0 & = \delta l_r^2 + 2 \mathbf{x}_{\rm in} \cdot \univec{r}
    \delta l_r + (r_{\rm in}^2 - r_\inter^2)\ .
    \\
    0 & = (\cos^2\theta_\inter-{z}_d^2) \delta
    l_\theta^2 + 
    2(\cos^2\theta_\inter \mathbf{x}_{\rm in}\cdot\univec{r}
    - z_i {z}_d) \delta l_\theta
    \\
    & + (\cos^2\theta_\inter - \cos^2 \theta_{\rm in} ) r_{\rm in}^2\ ;
    \\
    0 & = ({x}_d \tan\phi_\inter - {y}_d)
    \delta l_\phi + (x_{\rm in} \tan\phi_\inter - y_{\rm in})\ .
  \end{split}
\end{equation}
When solving these geometric equations for ray propagation,
the absence of real positive solutions to $\delta l$
indicates non-intersection with the corresponding coordinate
surface, allowing such cases to be safely disregarded. For
problems with reduced dimensionality or cells containing
degenerate surfaces (e.g., polar wedges in spherical grids),
the system \eqref{eq:method-rt-geometry} solves fewer
equations while maintaining three-dimensional
representations of position vector $\mathbf{x}_0$ and
direction vector $\hat{r}$.  \response{It is worth noting
  that multiple roots may emerge in curvilinear coordinate
  systems due to quadratic surface expressions
  (eqs.~\ref{eq:cyl-dl}, \ref{eq:sph-dl}), requiring
  explicit consideration. The exit point from each cell is
  determined by identifying the minimal positive solution
  $\delta l_{\min}$ among $\{\delta l_j\}$ corresponding to
  surface index $j$. Negative solutions indicate unphysical
  backward propagations, which have to be discarded. Note
  also that $\delta l = 0$ roots can naturally emerge when
  there are more than one intersection with cell surfaces,
  while these solutions indicating the point at which the
  ray arrives into the current cell should also be excluded
  from the ray propagation solution. The updated propagation
  origin becomes $\mathbf{x}_0 + \delta l_{\min}\hat{r}$,
  initiating iterative traversal until domain exit. An
  example of this ray-tracing approach on spherical polar
  grids is illustrated in Figure~\ref{fig:rad_tst_sph}, in
  which the direction of incoming rays are still parallel to
  the $x-y$ plane but tilted to $\theta = arctan(0.1)$ with
  the horizontal axis. Such tilt exhibits the correct
  propagation results, where multiple intersections between
  rays and their current single cell surfaces are
  involved. }

Photons within each ray are characterized by
energy-dependent counts $N_{\ph,\mathrm{init}}(h\nu)$,
assigned per directional emitter. \response{For instance,
  from an isotropic point sources}, the initialization
follows,
\begin{equation}
  \label{eq:method-rt-ph-num}
  N_{\ph,\mathrm{init}}(h\nu) = L_*(h\nu) \dfrac{\Delta
    \Omega}{4\pi}  \delta t, 
\end{equation}
where $L_*(h\nu)$ denotes spectral luminosity of the source,
$\Delta \Omega$ the ray's solid angle, and $\delta t$ the
timestep duration. Photon absorption within cells
incorporates mean free path $\lambda_{\abs}(h\nu)$ and
effective flux calculation via path-averaged integration:
\begin{equation}
  \label{eq:method-rad-equiv-flux}
  \begin{split}
    & \quad F_{\ph,\eff}(h\nu)\\
    &\simeq N_{\ph,\mathrm{in}}(h\nu) \left(
      \dfrac{\delta l_{\min}}{\delta V \delta t} \right)
      \left(\int_\mathrm{in}^\mathrm{out} \d s \right)^{-1}
    \\ 
    &\quad \times \int_\mathrm{in}^\mathrm{out} \d s
      \exp[-\delta l_{\min}/\lambda_{\abs}(h\nu)] \\ 
    &= N_{\ph,\mathrm{in}} \left( \dfrac{\delta l_{\min}}{\delta V
      \delta t} \right) \left\{ \dfrac{ 1 -
      \exp[-\delta l_{\min}/\lambda_{\abs}]}
      {\delta l_{\min}/\lambda_{\abs}} \right\}. 
  \end{split}
\end{equation}
Here $\delta V$ represents cell volume, with flux
asymptotically approaching
$N_{\ph,\mathrm{in}} \delta l_{\min}/(\delta V \delta t)$ in
optically thin regimes
($\delta l_{\min}/\lambda_{\abs} \rightarrow 0$) and scaling
inversely with optical depth at high absorption. Multi-ray
contributions sum across energy bins using
\eqref{eq:method-rad-equiv-flux}. Photon depletion follows
exponential attenuation:
\begin{equation}
  \label{eq:method-rt-abs}
  N_{\ph,\mathrm{out}}(h\nu) = N_{\ph,\mathrm{in}}(h\nu)
  \exp[-\delta l_{\min}/\lambda_{\abs}(h\nu)], 
\end{equation}
while accumulated column densities along propagation paths
provide critical inputs for subsequent thermochemical rate
calculations. \response{It is noted that, although the
  discussions are introduced assuming an isotropic source,
  application of the ray tracing system is not limited to
  this simplest case. While Figure~\ref{fig:rad_tst_sph}
  presents an example of parallel lights,
  \citet{2025arXiv251201283Y} have also verified the same
  module for ray tracing on the cases for extended sources
  (equivalent to $\sim 10^6$ sources distributed across the
  simulation domain). }

\subsection{Radiation-matter Interactions}
\label{sec:method-rad-reac}

In the \kratos{} computational framework, it is assumes
spatially uniform species concentrations within individual
cells by default. For monochromatic radiation traversing a
cell with photon luminosity $L$, the photon deposition rate
derives from Beer-Lambert attenuation as
$L[1 - \exp(-\sigma n \delta l)]$, where $\sigma$ denotes
reaction cross-section, $n$ reactant concentration, and
$\delta l$ optical path length. Normalization by reactant
count $n\delta V$ (with $\delta V$ as cell volume) yields
the reaction rate coefficient:
\begin{equation}
  \label{eq:photo-reaction-single}
  k \simeq \dfrac{L[1-\exp(-\sigma n \delta l)]}{n \delta V}
  = \left( \dfrac{L \sigma \delta l}{\delta V} \right)
  \left( \dfrac{1 - \exp(-\tau)}{\tau} \right), 
\end{equation}
where $\tau \equiv \sigma n \delta l$ represents the local
optical depth of absorption. This reduces to
$k = L\sigma\delta l/\delta V$ in optically thin regimes
($\tau \rightarrow 0$). Generalization to multi-reaction,
multi-ray systems employs the formalism with multiple
indices for reactant species $i$, radiation rays $j$, and
photon energy bins $k$,
\begin{equation}
  \label{eq:photo-reaction-all}
  \begin{split}
    k_i & = \sum_{j,k} k^{(0)}_{ijk} \left[\dfrac{ 1 - \exp(
          - \hat{\tau}_{jk})}{\hat{\tau}_{jk}} \right], \\ 
    k^{(0)}_{ijk}
        & \equiv \dfrac{L_{jk}\sigma_{ik}\delta l_j}{\delta
          V}, \quad \hat{\tau}_{jk} \equiv \sum_i n_i
          \sigma_{ik} \delta l_j, 
  \end{split}
\end{equation}
where $L_{jk}$ specifies photon flux per ray-energy pair,
$\sigma_{ik}$ energy-dependent cross-sections, and
$\delta l_j$ ray-specific path lengths. These rates directly
populate the matrix $\{B^i_{\;j}\}$ in
\eqref{eq:method-ode-thermochem}, with reactant depletion
($-k_i$) and product generation ($+k_i$) terms.

Temporal integration over timestep $\delta t$ quantifies
species concentration changes and energy-resolved photon
absorption:
\begin{equation}
  \label{eq:ray-absorption}
  \begin{split}
    \delta L_{jk}
    & = \sum_i\left( \dfrac{\sigma_{ik}\delta
      l_j}{\mathcal{N}_i} \int_t^{t+\delta t}  k_i n_i\ \d t
      \right), \\ 
    \mathcal{N}_i & \equiv \sum_{j,k}\sigma_{ik}\delta l_j,
  \end{split}
\end{equation}
where $\mathcal{N}_i$ normalizes cross-section-weighted path
lengths. The attenuated photon fluxes
$L_{jk} \leftarrow L_{jk} - \delta L_{jk}$ propagate
iteratively through downstream cells, coupling
\response{rays for photons} to chemical evolution via
cumulative optical depth updates.

\section{Code Verifications}
\label{sec:verify}

\response{ This section provides code verifications for the
  ray-tracing and thermochemistry modules combined. It is
  noted that the parallelization approach of the \kratos{}
  framework is built upon the heterogeneous architecture
  detailed in \citet{2025arXiv250102317W}, while the
  specific implementation and performance scaling of the
  ray-tracing radiation module are documented in details in
  \cite{2025arXiv251201283Y}. Within the context of the
  simulations presented in this work, which emphasize
  real-time non-equilibrium thermochemistry, the
  computational overhead of radiation transport is secondary
  to that of the stiff chemical ODE
  integration. Consequently, the comprehensive scaling tests
  and performance metrics provided in this manuscript
  primarily reflect the efficiency and throughput of the
  thermochemical solver, as it represents the dominant
  computational bottleneck in these coupled astrophysical
  applications.}

\subsection{Advection Tests}
\label{sec:verify-adv-spe}

\begin{figure}
  \centering
  \includegraphics[width=3.3in, keepaspectratio]
  {\figdir/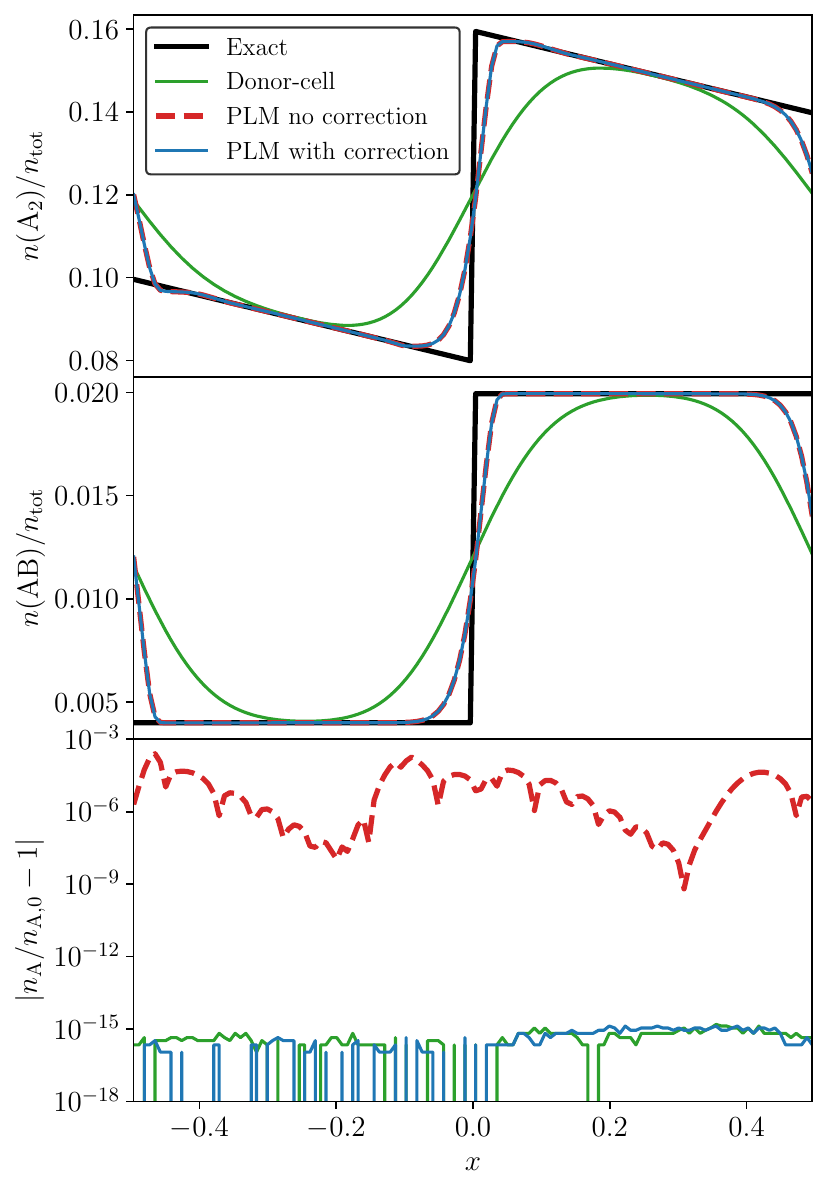}
  \caption{Test results of different advection schemes (see
    \S\ref{sec:verify-adv-spe}). The advection results (in
    number density ratios relative to $n_{\rm tot}$) using
    different algorithms are presented in the upper two
    panels at $t=1$ for two representative species, compared
    to the exact solution in heavy solid lines. Note that
    the lines for PLM with and without stoichiometric
    corrections overlap each other. The bottom panel shows
    the violation of elemental abundance conservation in
    comparison to the exact solutions.}
  \label{fig:adv-test}
\end{figure}

To evaluate the conservation properties of the advection
algorithms described in \S\ref{sec:method-cons-advec},
\response{and specifically to demonstrate the necessity of
  implementing conservative higher-order reconstruction for
  chemical species (rather than donor-cell methods) in
  \kratos{} }, a one-dimensional test case is implemented
within the spatial domain $x \in [-0.5, 0.5]$, discretized
into $128$ uniform zones with periodic boundary
conditions. The fluid maintains constant mass density
($\rho = 1$) and pressure ($p = 1$) in code units, while
advecting left-to-right at velocity $v = 1$. Four passive
chemical species (\chem{A}, \chem{A_2}, \chem{AB}, \chem{B})
track two conserved elements (A and B), with initial
abundance profiles defined as:
\begin{equation}
  \begin{split}
    n_\chem{A} & = 0.49 \Theta(-x) + 0.05\Theta(x) + 0.2x, \\
    n_\chem{A_2} & = 0.20 \Theta(-x) + 0.40\Theta(x) - 0.1x, \\
    n_\chem{AB} & = 0.01 \Theta(-x) + 0.05\Theta(x), \\
    n_\chem{B} & = 0.09 \Theta(-x) + 0.05\Theta(x),
  \end{split}
\end{equation}
where $\Theta(x)$ denotes the Heaviside step function. The
system evolves for $t=1$, which should theoretically recover
initial conditions due to periodicity.

Figure~\ref{fig:adv-test} reveals two key
observations. First, the piecewise linear method (PLM)
reconstruction scheme significantly reduces numerical
diffusion compared to donor-cell approaches, regardless of
conservation corrections. Second, uncorrected PLM introduces
substantial element A conservation errors ($\sim 10^{-3}$
per $\sim 300$ step), which would grow to $\sim 10^{-1}$
after $\sim 10^5$ timesteps, being even more catastrophic
for longer-term thermochemical simulations. In contrast,
both donor-cell and corrected PLM schemes maintain species
conservation within machine precision ($\lesssim 10^{-15}$
relative error), demonstrating the necessity of the flux
correction mechanism described in
\S\ref{sec:method-cons-advec}.


\subsection{Single-point Thermochemistry Tests}
\label{sec:verify-0d}

\begin{figure}
  \centering
  \includegraphics[width=3.5in, keepaspectratio]
  {\figdir/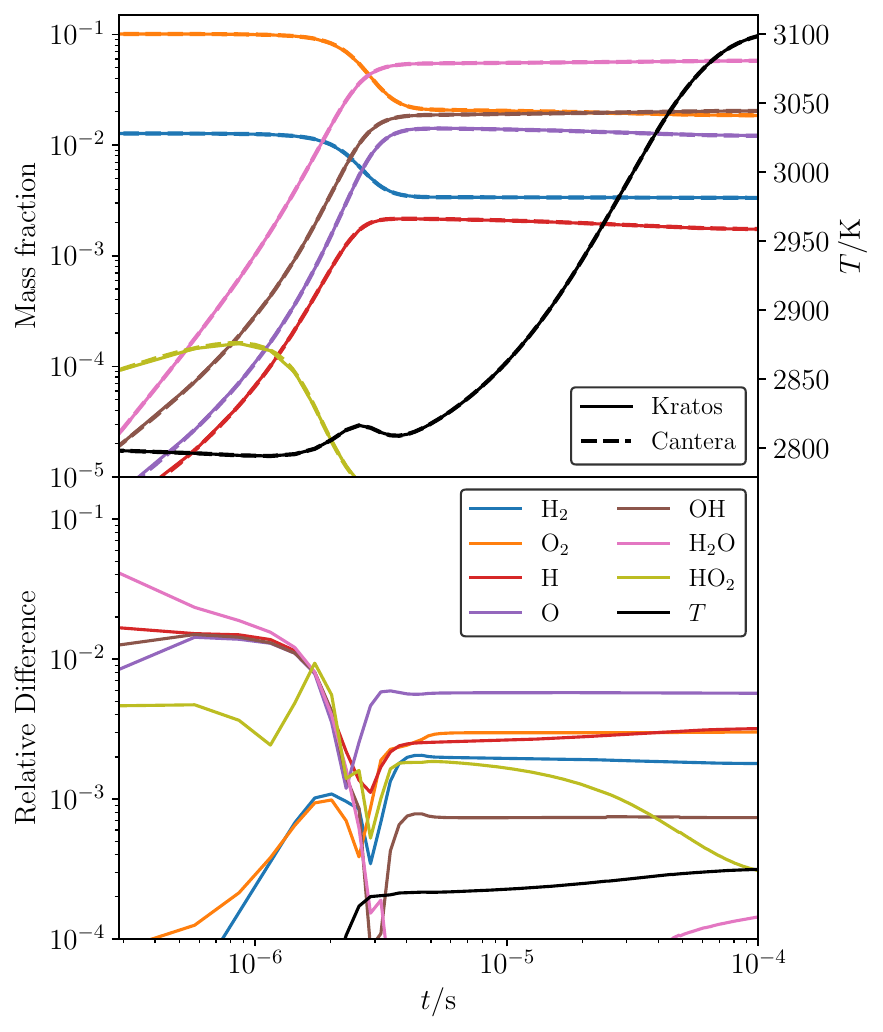} 
  \caption{Thermochemical reaction tests for the burning of
    molecular hydrogen with oxygen (\S\ref{sec:verify-0d}),
    comparing the \kratos{} mixed-precision results (solid
    lines) with the \cantera{} ones (dashed lines). The
    upper panel exhibits the evolution of different
    variables (distinguished by line colors), including the
    mass fraction of chemical species (the left ordinate, in
    logarithm scale) and the temperature (the right
    ordinate, in linear scale), while the lower panel shows
    the relative differences of the mass fractions and the
    gas temperature. }
  \label{fig:chem-uni}
\end{figure}

To validate the thermochemical solver in the absence of
hydrodynamic effects, single-point ignition tests were
conducted for a stoichiometric mixture of molecular hydrogen
(\chem{H_2}) and molecular oxygen (\chem{O_2}) diluted in
argon (Ar). The simulations employed the GRI-Mech 3.0
chemical kinetic mechanism \citep{grimech}, using the
submechanism subtracted by \cantera{}
\citep{2022zndo...6387882G} prioritizing high-temperature
combustion processes. The submechanism is a
combustion-focused network comprising 12 species (including
radicals and intermediates such as \chem{H}, \chem{O},
\chem{OH}, and \chem{HO_2}) and 116 elementary reactions
(including inverse reactions). The initial conditions were
defined as a molar composition of 20\% \chem{H_2}, 10\%
\chem{O_2}, and 70\% Ar, preheated to $T = 2800~\K$ to
accelerate ignition. The evolution track is simulated using
both the \kratos{} framework and the benchmark tool
\cantera{}, with a focus on tracking species concentrations
and temperature until thermochemical equilibrium was
achieved.

A key consideration in high-temperature combustion
simulations is the accurate representation of heat capacity
($c_V$) variations. At temperatures exceeding $\sim 10^3~\K$
, vibrational modes in polyatomic molecules (e.g.,
\chem{H_2O}, \chem{O_2}) become thermally activated, leading
to a nonlinear rise in $c_V$ with increasing
temperature. The \kratos{} solver therefor includes an
option to incorporate this physics through NASA polynomial
formulations \citep{MBNMH2010NASA} using updated parameters
adopted by \cantera{} \citep{2022zndo...6387882G}, which
express $c_V$ into piecewise sixth-order polynomial
functions across discrete temperature intervals (typically
$300-1000~\K$ and $1000-3500~\K$). These polynomials are
derived from spectroscopic data and statistical
thermodynamics, enabling precise calculations of enthalpy
and entropy. To quantify the impact of this model, a control
simulation assuming constant heat capacity (frozen at the
$300~\K$ values) was performed, resulting in a $\sim 15\%$
overprediction of equilibrium temperature due to the
underestimation of heat capacities. This error arises from
neglecting the energy partitioned into vibrational modes,
which effectively increases the system's thermal
inertia. The integration of NASA polynomials into
equation-of-state class template of \kratos{} ensures that
such temperature-dependent effects are rigorously captured,
a necessity for predictive combustion modeling.

Figure~\ref{fig:chem-uni} demonstrates high consistency
between the \kratos{} and \cantera{} results across the
combustion timeline. During the ignition phase
($t\lesssim 10^{-6}~\s$), minor discrepancies in trace
species (e.g., \chem{H}, \chem{O}, \chem{OH}) with mass
fractions below $10^{-3}$ were observed, exhibiting relative
differences on the order of $\sim 10^{-2}$. These deviations
are attributed to numerical sensitivities in resolving fast
radical-driven reactions during the induction
period. However, as the system approached equilibrium, the
mass fractions of major species--\chem{H_2O} (primary
product), unburned \chem{H_2}, and residual
\chem{O_2}--converged to near-identical values in both
solvers, with discrepancies limited to $\lesssim
10^{-3}$. Notably, the final equilibrium temperature
computed by \kratos{} differed from \cantera{} by only
$10^{-4}$ relative error, emphasizing the precision of the
thermodynamic and ray-tracing models implemented in
\kratos{} even using mixed-precision methods. The close
agreement between \kratos{} and \cantera{} validates the
former's implementation of chemical kinetics, thermodynamic
properties, and solver algorithms. The convergence of major
species and temperature to near-machine precision at
equilibrium also exhibits the robustness of \kratos{}
semi-implicit solver and its adaptive substep control
scheme.


\subsection{Str\"omgren Sphere Tests}
\label{sec:verify-strom}

\begin{figure*}
  \centering
  \includegraphics[width=4in, keepaspectratio]
  {\figdir/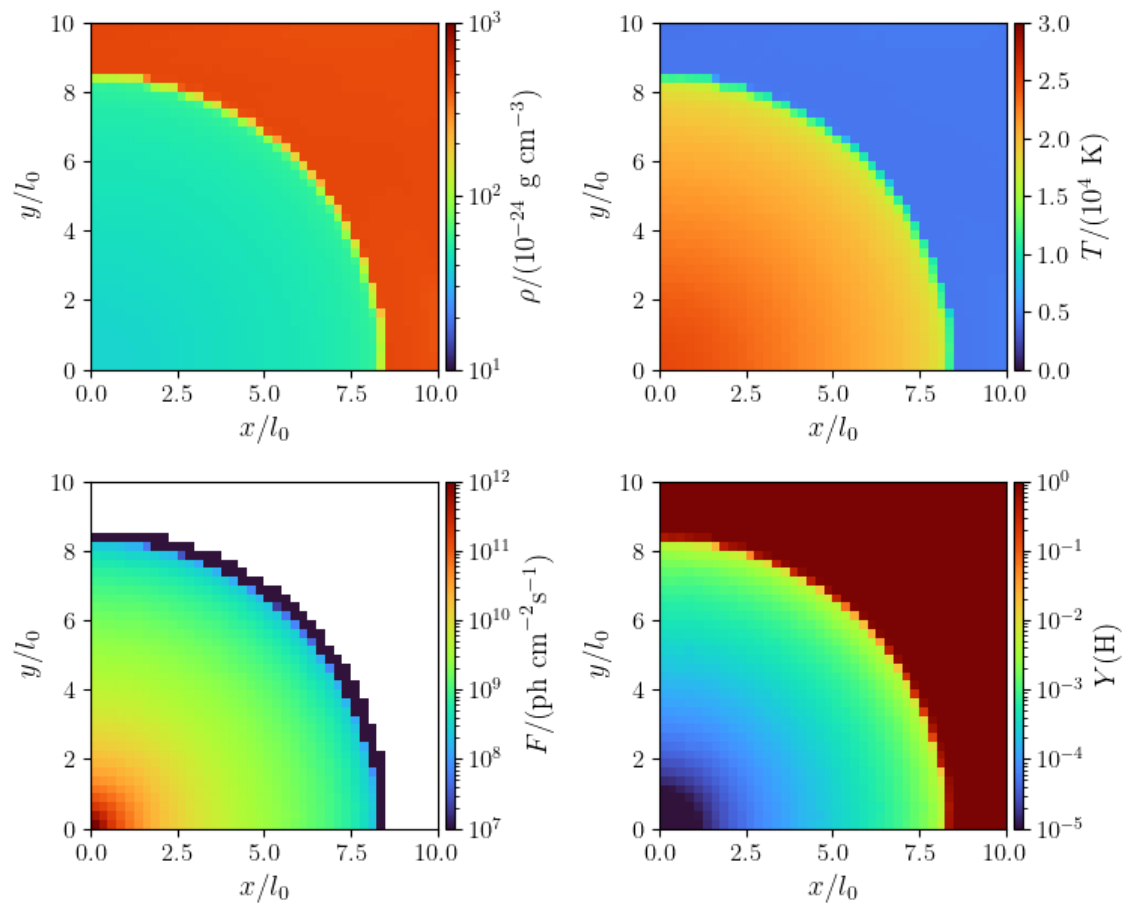} 
  \includegraphics[width=2.9in, keepaspectratio]
  {\figdir/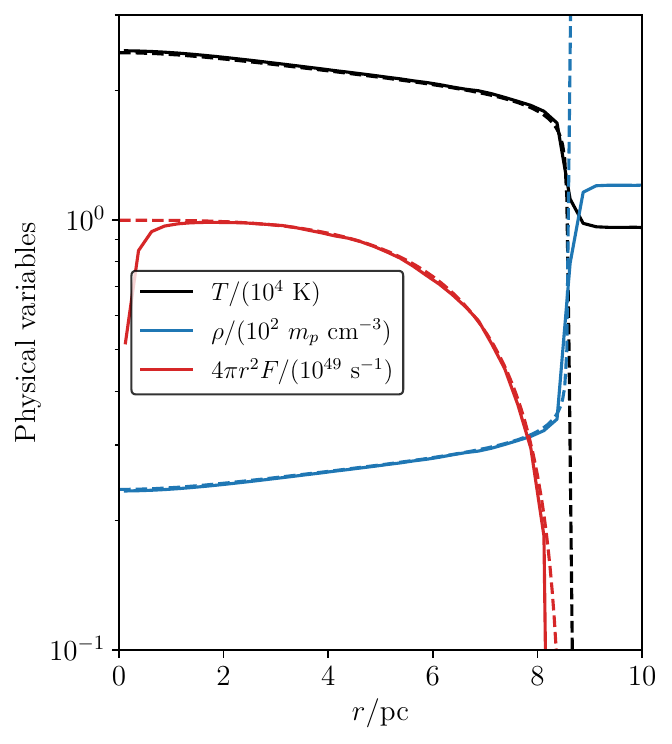} 
  \caption{Hydrostatic profiles of the Str\"omgren sphere
    test using the fiducial set of parameters (see
    \S\ref{sec:verify-strom}). The left four colormap
    panels, marked with the code unit lenght $l_0 = \pc$,
    show the slices at $z=0$ for mass density $\rho$,
    temperature $T$, ionizing photon flux $F$, and the mass
    fraction of neutral hydrogen $Y(\chem{H})$. The right
    panel illustrates the profiles along the
    radial line along the $z$-axis,
    comparing the results by \kratos{} (solid lines) with
    the semi-analytic solutions (dashed lines). }
  \label{fig:stroem-profile}
\end{figure*}


Str\"omgren sphere tests serve as a standard for validating
numerical simulations of thermochemical reactions and
\response{radiation calculations} in astrophysical
scenarios. Such tests model the ionized regions around
luminous sources of ionizing photons embedded in a neutral
medium, providing a well-understood benchmark for comparing
numerical methods against semi-analytic solutions. This work
employs the Str\"omgren sphere test to evaluate the accuracy
of the radiation-hydrodynamics framework, focusing on the
interplay between photoionization, collisional processes,
and radiative recombination in a simplified hydrogen-only
chemistry network that only consist of atomic hydrogen
(\chem{H}), hydrogen ions (\chem{H^+}), and free electrons
($e^-$).
\begin{equation}
  \label{eq:strom-chem}  
  \begin{split}
    & \chem{H} + h \nu \rightarrow \chem{H^+} + e^- \ ;
      \quad
      \chem{H} + e^- \rightarrow \chem{H^+} + 2 e^- \ ;\\
    & \chem{H^+} + e^- \rightarrow \chem{H} + h \nu \ .\\
  \end{split}  
\end{equation}
At equilibrium, the thermal pressure inside the Str\"omgren
sphere equals the ambient pressure of the surrounding
neutral medium, given by the ambient hydrogen atom number
density $n_{\rm amb}$ and temperature $T_{\rm amb}$. The
internal profiles of Str\"omgren spheres are also
constrained by ionization, recombination, and
thermodynamical conditions,
\begin{equation}
  \label{eq:strom-equil}
  \begin{split}
    & F = \dfrac{\Phi}{4\pi r^2} \e^{-\tau}\ ,\ \dfrac{\d
      \tau}{\d r} = n_\H \sigma\ ,\\    
    & 0 = n_\H (1 + x_e) \kb T - n_{\rm amb}\kb
      T_{\rm amb}\ ,\\
    & 0 = \alpha n_\H^2 x_e^2 - k_{\rm ci}n_\H^2 x_e(1-x_e)
      - F \sigma n_\H (1-x_e) \ , \\
    & 0 = \lambda_{\rm rec} \alpha n_\H^2 x_e^2 + I_\H
      k_{\rm ci}n_\H^2 x_e(1-x_e) \\
    & \qquad - ( h\nu - I_\H) F \sigma n_\H (1-x_e) \ .\\
  \end{split}
\end{equation}
Here, $F$ is the ionizing photon flux, $\Phi$ is the
luminosity of ionizing photons, $\tau$ is the absorption
optical depth, $\sigma$ is the photoionization
cross-section, $n_\H$ is the total number density of
hydrogen nuclei (including \chem{H} and \chem{H^+}), $h\nu$
is the photon energy (assuming monochromatic radiation), and
$I_\H = 13.6~\eV$ is the ionization energy of hydrogen. The
thermochemistry parameters are functions of temperature, including
$k_{\rm ci}$ as the collisional ionization rate coefficient,
$\alpha$ as the case-B recombination coefficient, and
$\lambda_{\rm rec}$ as the cooling energy per recombination
reaction,
\begin{equation}
  \label{eq:thermo-strom}
  \begin{split}
    & k_{\rm ci} = 1.05\times 10^{-9}~\cm^3~\s^{-1}\
    \left( \dfrac{T}{300~\K} \right)^{1/2}
    \e^{-I_\H/\kb T} ,\\
    & \alpha =  3.5\times 10^{-12}~\cm^3~\s^{-1}\
    \left( \dfrac{T}{300~\K} \right)^{-0.75} , \\
    & \lambda_{\rm rec} = \left[0.684 - 0.0416
      \ln\left( \dfrac{T}{10^4~\K}\right) \right] \kb T\ .\\
  \end{split}
\end{equation}
adopting the data and expressions in \citet{UMIST2013} and
\citet{DraineBook}. Eqs.~\eqref{eq:strom-equil} with
\eqref{eq:thermo-strom} can be solved semi-analytically to
obtain the hydrodynamics, radiation, and thermochemical
profiles, by integrating from the center ($r = 0$) to the
location that the absorption optical depth reaches
$\tau = 10^2$ (other terminal $\tau$ values are tested to
verify that the results are not affected).

\begin{figure}
  \centering
  \includegraphics[width=3.35in, keepaspectratio]
  {\figdir/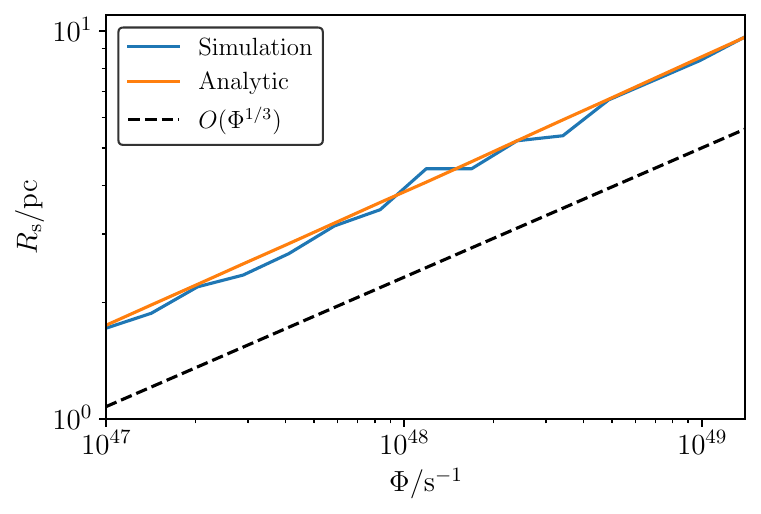} 
  \caption{Radii of Str\"omgren spheres ($R_{\rm s}$) under
    different ionizing photo luminosities ($\Phi$),
    comparing the simulation results to the analytic
    solutions. The $R_{\rm s}\propto \Phi^{1/3}$ power-law
    is shown in a heavy dashed line for reference. }
  \label{fig:stroem-radius}
\end{figure}

Using the radiation and thermochemical module described in
\S\ref{sec:method-chem} and \S\ref{sec:method-rt}, a
fiducial Str\"omgren sphere with parameters
$\Phi = 10^{49}~\s^{-1}$, $T_{\rm amb} = 10^4~\K$, and
$n_{\rm amb} = 130~\cm^{-3}$ is solved by a \kratos{}. The
simulation locates the radiation source at the origin point
of a mesh with $64^3$ zones, covering the $[0,\ 16~\pc]^3$
spatial region (thus $\Delta x = 1/4~\pc$), and is evolved
through the steady state with constant pressure at
boundaries given by $n_{\rm amb}$ and $T_{\rm amb}$. Note
that a relatively low resolution is adopted intentionally to
demostrate the accuracy and robustness of the system under
limited resolution.  Figure~\ref{fig:stroem-profile}
exhibits excellent agreement between the numerical results
and semi-analytic profiles within the ionized region, with
relative deviations typically below $\sim
5\%$. Discrepancies emerge in the innermost cells
($r < 1~\pc$), where the radiation flux is underestimated by
up to $\sim 50\%$. This deviation is confirmed to be arising
from the Cartesian grid's limited ability to resolve
spherical symmetry at low radii, where the misalignment of
the grid with the intrinsic geometry becomes
significant. Despite this, the outer regions, where
ionization and thermal structures are most dynamically
relevant, show robust convergence, validating the numerical
treatment of ionization and radiation co-evolved with
hydrodynamics.

Figure~\ref{fig:stroem-radius} presents the simulated
Str\"omgren radii $R_{\rm s}$, which agree with
semi-analytic predictions to within $\sim 5\%$ across two
orders of magnitude in ionizing photon luminosity $\Phi$
(from $10^{47}~\s^{-1}$ through $10^{49}~\s^{-1}$). This
consistency also verifies the ability of \kratos{} framework
to accurately capture the $R_{\rm s}\proptosim \Phi^{1/3}$
scaling of the Str\"omgren radius, a fundamental relation in
HII region physics. The minor deviations at low $\Phi$ also
stem from the geometry, when the sphere radii are small and
the discreteness of the mesh cells cause the step growth
when $\Phi$ increases. The Str\"omgren sphere test confirms
that the \kratos{} framework accurately reproduces key
physical processes in ionized plasmas within the sphere,
including chemical reactions (ionization balance),
thermodynamics, and \response{radiation}.

\subsection{Detonation Tests}
\label{sec:verify-detonation}

\begin{figure}
  \centering
  \includegraphics[width=3.35in, keepaspectratio]
  {\figdir/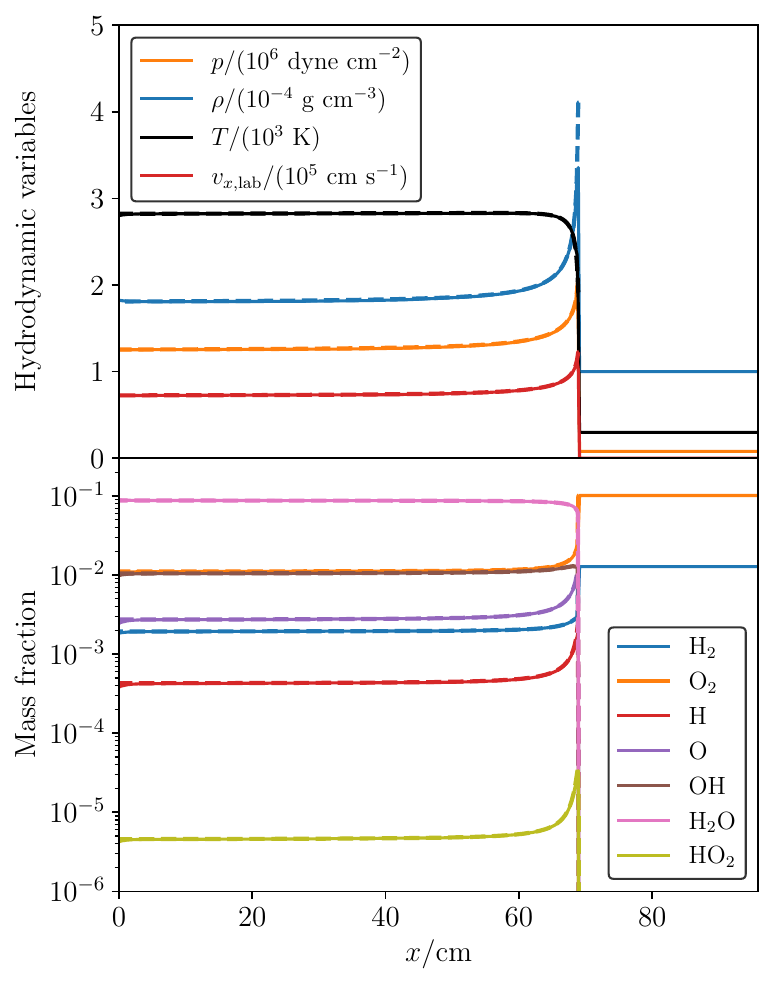} 
  \caption{Hydrodynamic profiles (upper panel) and mass
    fractions of chemical species (lower panel) of the
    fiducial detonation test (see
    \S\ref{sec:verify-detonation}), comparing the \kratos{}
    simulation results (solid lines) with the semi-analytic
    results yielded by SDT (dashed lines). Note that the
    presented velocity is measured in the lab-frame. }
  \label{fig:det-profile}
\end{figure}


\begin{figure}[t]
  \centering
  \includegraphics[width=3.35in, keepaspectratio]
  {\figdir/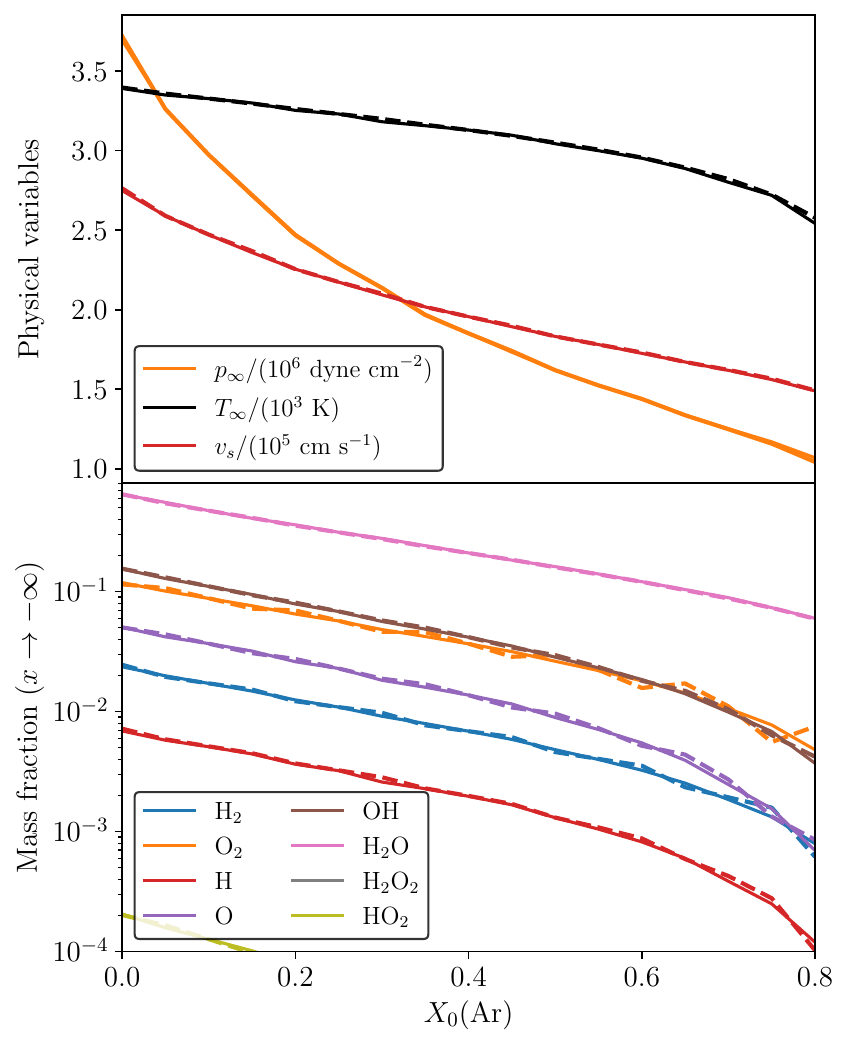} 
  \caption{Hydrodynamic quanitites (upper panel) and mass
    fractions of chemical species (lower panel) with
    different upstream molar fractions of argon
    [$X_0(\chem{Ar})$], showing the detonation shock speed
    $v_s$, and physical quantities at large distance in the
    detonation wave downstream (marked as
    $x\rightarrow -\infty$). The \kratos{} simulations
    (solid lines) are compared to the SDT results (dashed
    lines). Note that the mass fractions of low-abundance
    species at high $X_0(Ar)$ are slightly unstable in the
    SDT outputs. }
  \label{fig:det-compare}
\end{figure}

Detonation flows represent one of the most rigorous
benchmarks for reacting flow simulation codes, as they
involve complex interactions between hydrodynamics, chemical
reactions \response{with multiple chemical species, and
  thermodynamics associated with these reactions}.  A
detailed validation is carried out for the \kratos{}
simulation framework simulating steady-state detonations,
comparing numerical results against semi-analytic solutions
computed with the Shock and Detonation Toolbox (SDT
hereafter; \citealt{MBNMH2010NASA}) constructed on the top
of \cantera{}. These tests not only assess the accuracy of
the thermochemical network but also evaluate the code
ability to resolve shock propagation, and the chemistry at
and in the downstream of detonation shocks in
multi-dimensional configurations.  Ideally, each set of
upstream conditions corresponds to a unique set of
downstream conditions, and can be described by semi-analytic
solutions. To develop into this steady-state detonation
described exactly by semi-analytic solutions, however, the
shock wave has to propagate through a very long spatial
range to converge to the steady state. Therefore, the
simulation condition in these detonation tests are setup in
the shock frame, while the upstream gas is fed into the
system from the boundary on the right-hand side.

\subsubsection{Comparisons with the SDT Benchmark}

The detonation tests are configured in one spatial dimension
(with additional verification in higher dimensions) using
the thermochemical network described in
\S\ref{sec:verify-0d}. The simulations are initialized in
the shock frame, where upstream gas is fed into the system
from the right-hand boundary, while the left-hand boundary
adopts outflow conditions. This setup ensures that the shock
wave propagates through a sufficiently long spatial distance
to achieve steady-state conditions, which are critical for
meaningful comparisons with semi-analytic solutions.  For
the fiducial simulation carried out on a $(x/\cm)\in[0, 96]$
domain covered by $512$ simulation zones along the
$x$-axis, the upstream gas consists of 20\% \chem{H_2},
10\% \chem{O_2}, and 70\% Ar by molar fraction, with a
temperature $T_0 = 298~\K$ and mass density
$\rho_0=10^{-4}~\g~\cm^{-3}$. At $t = 0$, the region with
$x < 76.8~\cm$ is set uniformly to the post-shock values at
$x\rightarrow -\infty$ for the density, temperature, and
chemical abundances given by SDT, in order to trigger the
shock, the chemical reactions, and to shape the detonation
eventually. This approach ensures consistency between the
numerical and semi-analytic frameworks, minimizing
discrepancies arising from mismatched initial states.

Figure~\ref{fig:det-compare} showst the comparison between
\kratos{} simulation results and SDT semi-analytic
solutions, revealing excellent agreement across all
hydrodynamic and thermochemical profiles. Relative errors in
downstream variables (e.g., density, temperature, velocity)
are consistently at the order of $\sim 0.1\%$, demonstrating
the high fidelity of the numerical method. Most notably, the
detonation speed--a critical parameter for validating shock
propagation--deviates from the semi-analytic solution by no
more than $0.3\%$. This precision underscores the
framework's ability to resolve the intricate coupling
between hydrodynamics and chemistry in detonation flows.

\subsubsection{Accuracy under Various Conditions}

To further validate the robustness of the framework, more
tests are conducted to vary upstream Ar abundance
systematically while maintaining stoichiometric ratios of
\chem{H_2} and \chem{O_2}. The downstream hydrodynamic
variables (e.g., pressure, temperature) at sufficient
distance from the detonation front (marked as
$x\rightarrow -\infty$) agree with SDT results within
$\sim 1\%$, confirming the predictive capability of
\kratos{} across a range of upstream conditions.  Minor
discrepancies emerge in the mass fractions of chemical
species with low abundance (e.g., \chem{OH}, \chem{H_2O}),
which fluctuate by $\sim 10\%$ when computed using SDT. This
behavior is attributed to the reliance of \cantera{} (as the
backend of SDT) on operator-splitting methods, where
thermodynamics and chemistry are solved in separate
sub-steps. In comparison, \kratos{} employs a simultaneous
solution approach (\S\ref{sec:method-algo-cpu}), which
reduces numerical instabilities and produces smoother, more
physically consistent results. This illustrates a key
advantage of the \kratos{} framework in resolving complex
reacting flows, particularly in regimes where minor species
play significant roles.

\subsubsection{Speed Tests in 3D}

The detonation test series serves dual purposes--validating
thermochemical accuracy and quantifying computational
efficiency in \kratos{}. By extending the fiducial
detonation setup to plane-parallel 3D geometry, the
performance across various heterogeneous architectures is
benchmarked and presented in
Table~\label{table:spd-det}. Contemporary GPUs demonstrate
\response{high computational throughput}, with NVIDIA RTX
4090 achieving $\sim 1.7 \times 10^7{\rm\ cell\ s}^{-1}$
initial computation rates as shown in
Table~\ref{table:spd-det}, \response{along with a few other
  typical architectures}. This outperforms CPU-based
\cantera{} implementations by orders of magnitude
($\sim 10^6{\rm\ cell\ s}^{-1}$ with 8-core parallelization,
and $\sim 10^5{\rm\ cell\ s}^{-1}$ on a single
core). \response{Comparisons with more CPU cores have also
  been conducted, yet the basic scaling remain the
  same. Note also that, on a typical contemporary multi-GPU
  server, one GPU usually corresponds to 8 or 16 CPU
  cores). It is also noted that, using identical codebases,
  the thermochemical module has been derived into
  thermonuclear reaction module, which have also been
  calibrated and tested for performance against artificial
  neural network methods in \citet{2025ApJ...990..105Z}.  }

Compared to pure hydrodynamic simulations elaborated in
\citet{2025arXiv250102317W}, thermochemical integration
introduces $\sim 10^2$ slowdown due to stiff ODE solving
with semi-implicit methods, as the sub-step controllers
dynamically adjust temporal resolution, increasing iteration
counts during chemically active phases.  Despite this
baseline cost, thermochemistry exhibits favorable strong
scaling characteristics. \response{Mixed-precision
  implementations, together with the MPI-based inter-process
  data transfer schemes detailed in
  \citet{2025arXiv250102317W}, achieve $\sim 84\%$ parallel
  efficiency across 8 GPUs,} contrasted with $\sim 50\%$
scaling for hydrodynamic-only calculations on equivalent
hardware \citep[see also][] {2025arXiv250102317W}. This
increase arises from thermochemistry's embarrassingly
parallel workload distribution, where most of the computing
cycles remain local to individual cells. The remaining
overhead still arises from global \response{ray-tracing}
updates and MPI boundary exchanges.


\begin{deluxetable}{lccc}
  \tablecolumns{4} 
  \tabletypesize{\scriptsize}
  \tablecaption{Performance measurements based on the 3D
    detonation tests.
    \label{table:spd-det}
  } \tablehead{ \colhead{Programming Models} &
    \multicolumn{3}{c}{Computing Speed with Precision}
    \\
    \colhead{and Devices} &
    \multicolumn{3}{c}{($10^6{\rm\ cell\ s}^{-1}$)}
    \\
    \cline{2-4} \colhead{} & \colhead{Single} &
    \colhead{Mixed} & \colhead{Double} } \startdata
  HIP-CPU${}^*$ & & & \\
  AMD Ryzen 5800X${}^\dagger$ & & & 0.19 \\
  Qualcomm Snapdragon 888${}^{**}$ & & & $0.0089$ \\
  \\
  NVIDIA CUDA & & & \\
  RTX 3090   & 6.9  & 6.5  & 0.97 \\
  RTX 4090   & 21.6  & 17.1  & 2.4 \\
  RTX 4090$\ \times 8^\ddagger$ &  114.9 & 112.4  & 16.3  \\
  Tesla A100 & 9.3 & 9.1  & 5.5 \\
  \\
  AMD HIP & & & \\
  7900XTX & 4.5 & 4.6 & $-^{***}$ \\
  MI100   & 1.52 & 1.56 & $-^{***}$ \\
  \enddata \tablecomments{ Presenting the average over
    $10^2$ steps. All tests cases are in 3D. Detailed setups
    see \S\ref{sec:verify-detonation}.  \\
    *: Only double precision results are concerned, as
    modern CPUs have almost the same single
    and double precision computing speeds.  \\
    $\dagger$: Utilizing all 8 physical cores.\\
    $\ddagger$: Using 8 GPU cards on the same computing
    node, with the same simulation setup (i.e. showing the
    strong scaling). \\
    **: Using termux (\url{https://termux.dev}) on Android
    operating system, utilizing one major physical
    core. Compile-time optimization are turned {\it off}
    because of the software restrictions of \code{TBB} on
    ARM CPUs. \\
    ***: Failed to launch on the AMD HIP model with full
    double precision, due to the lack of several double
    precision functions. }
\end{deluxetable}

\section{Discussions and Summary}
\label{sec:summary}

The thermochemistry and ray tracing modules presented in
this paper, based on the \kratos{} framework
\citep{2025ApJS..277...63W}, provides a comprehensive and
efficient solution for simulating complex reacting flows in
astrophysical and other scenarios.
By integrating consistent microphysics calculations with
hydrodynamics, the \kratos{} framework overcomes limitations
of previous approaches, such as computational inefficiencies
and inconsistencies between thermochemistry and fluid
dynamics.

\kratos{} allows full degrees of freedom to include
thermochemical processes in the runtime by users. Key
innovations incorporated in the thermochemistry and ray
tracing modules of \kratos{} involve the
heterogeneous-optimized algorithms of thermochemical
evolution, ray-tracing, and chemical species advection,
which enable high-performance simulations while ensuring
element conservation and numerical stability. The
stoichiometry-compatible scheme for the reconstruction of
chemical species is applicable to higher order schemes of
reconstruction that may involve non-linear functions, which
removes the constraint of constant relative abundances of
elements in other methods such as CMA, and accelerates the
calculations by avoiding matrix inversions. The algorithms
for solving thermochemical ODEs through parallel LU
decomposition on SIMT devices also maximizes the efficiency
to conduct reacting flow simulations on heterogeneous
devices, especially GPUs.

The robustness and accuracy of \kratos{} are demonstrated
through a series of rigorous tests, including chemical
species advection, single-point thermochemical evolution,
Str\"omgren sphere simulations, and detonation flow
benchmarks. These tests validate the ability of \kratos{} to
maintain elemental conservation, resolve complex
hydrodynamic structures, and accurately model ionization
fronts and combustion processes, with results closely
matching benchmarks like against analytical results and
semi-analytic benchmarks given by \cantera{} and SDT. In
addition, the computing speed of \kratos{} with
thermochemistry also outperforms CPU-based methods (used in
e.g. \cantera{}), scaled at $\sim 10^2$ CPU cores using one
contemporary GPU.  The \kratos{} framework thus exhibits its
versatility in addressing diverse astrophysical phenomena
that involve thermochemistry with radiation as the
underlying driving mechanisms.

Looking ahead, several directions for future work are
envisioned to further enhance the capabilities of \kratos{}
framework. The ray-tracing scheme described in
\S\ref{sec:method-rt} can be extended without too much
efforts to solve the \response{radiative transfer problems
  with scattering, which enables a much broader range of
  applications dealing with radiation-matter
  interactions. This module is also posible to be developed
  into a stand-alone distribution of \kratos{} that deals
  with scattering of polarized photons, which can be adopted
  by the exploration and interpretation of observations with
  polarization information \citep{2025arXiv251201283Y}. }
Machine learning techniques could be explored to accelerate
thermochemical rate evaluations or optimize solver
performance. For example, the DeepODE solver
\citep{deepode2025} is expected to futher accelerates the
reacting flow computations in \kratos{}. The current
attempts of integrating DeepODE into \kratos{} shows that
the machine learning solver performs better than the
``traditional'' sovler (e.g. \S\ref{sec:method-algo-cpu})
when the stiffness of the thermochemical ODEs are high,
while the traditional solver performs better than the
machine learning methods in the low-stiffness secular
evolution \citep{2025ApJ...990..105Z}. Extending support for
nuclear reaction networks, complicated equations of state,
and relativistic hydrodynamics would enable the studies of
extreme physical and astrophysical phenomena, which has been
already composed for the purpose of supernovae and compact
object mergers, and will be described in separate papers
\citep[see also][]{2025ApJ...990..105Z}. In the
``downstream'' of the thermochemical modules, the
integration with magnetic fields and non-ideal MHD effects
would extend the applicability to magnetized jets and
accretion disks, similar to the toolchain adopted in \citet
{2019ApJ...874...90W, 2024ApJ...973...37Y}.  From the
prospective of software engineering, improving user
accessibility through \code{Python} interfaces will
facilitate broader applications, and also allows for further
interactions and integrations of other community-driven
codes and modules.

\bigskip

Code availability: As \kratos{} is still being developed
actively, the author will only provide the code upon
requests and collaborations at this moment. While several
important modules that are alerady mature have already been
adopted and made public along with
\citet{2025ApJS..276...40W}, a more complete version of
\kratos{} will be available publicly after further and
deeper debugs are accomplished.

\bigskip

\noindent L. Wang acknowledges the support in computing
resources provided by the Kavli Institute for Astronomy and
Astrophysics in Peking University. The author thanks the
colleagues for helpful discussions (alphabetical order):
Xue-Ning Bai, Renyue Cen, Jeremy Goodman, Xiao Hu, Mordecai
Mac-Low, Kengo Tomida, Haifeng Yang, Tianhan Zhang, and Yao
Zhou, for helpful discussions and useful
suggestions. Special thanks to Tianhan Zhang for the
discussions on the physics of thermochemical detonation.

\bibliographystyle{aasjournal}
\bibliography{method}

\end{document}